\documentclass[12pt]{article}

\usepackage[top=30truemm,bottom=30truemm,left=25truemm,right=25truemm]{geometry}
\usepackage[superscript]{cite}
\usepackage{authblk}
\usepackage{amsmath} 
\usepackage{amssymb}
\usepackage{graphicx}
\usepackage{amsfonts}
\usepackage{dcolumn}% Align table columns on decimal point
\usepackage{bm}% bold math
\usepackage[pagewise,displaymath, mathlines]{lineno}% Enable numbering of text and display math
\title{Detection of the phase shift of an alternating-current magnetic field 
by quantum sensing with multiple-pulse decoupling sequences}
\author[1,*]{Toyofumi Ishikawa}
\author[1]{Akio Yoshizawa}
\author[1]{Yasunori Mawatari}
\author[1, 2]{Satoshi Kashiwaya}
\author[1]{Hideyuki Watanabe}

\affil[1]{Electronics and Photonics Research Institute,
National Institute of Advanced Industrial Science and Technology, Tsukuba, 305-8565, Japan}
\affil[2]{Department of Applied Physics, Nagoya University, Nagoya, 464-8603, Japan}

\affil[*]{toyo-ishikawa@aist.go.jp}

\begin{document}
%\pagewiselinenumbers

\maketitle

\begin{abstract}
Magnetometry utilizing a spin qubit in a solid state possesses high sensitivity.
In particular, a magnetic sensor with a high spatial resolution can be achieved with
the electron-spin states of a nitrogen vacancy (NV) center in diamond.
In this study, we demonstrated
that NV quantum sensing based on multiple-pulse decoupling sequences 
can sensitively measure not only the amplitude but also the phase shift of an alternating-current (AC) magnetic field.
In the AC magnetometry based on decoupling sequences,
the maximum phase accumulation of the NV spin due to an AC field can be generally obtained
when the $\pi$-pulse period in the sequences matches the half time period of the field
and the relative phase difference between the sequences and the field is zero.
By contrast, the NV quantum sensor acquires no phase accumulation if the relative phase difference is $\pi/2$.
Thus, this phase-accumulation condition does not have any advantage for the magnetometry.
However, we revealed that the non-phase-accumulation condition is available for detecting a very small phase shift
of an AC field from its initial phase.
This finding is expected to provide a guide for realizing sensitive measurement of a complex AC magnetic field
in micrometer and nanometer scales.
\end{abstract}

\flushbottom
\thispagestyle{empty}
\newpage

Over the past few decades, the demand for sensitive magnetic sensors
and magnetometers has been growing not only in fundamental investigations of magnetism~\cite{doi:10.1063/1.3491215}
and spin dynamics~\cite{doi:10.1063/1.2943282, RevModPhys.81.1495}
but also in a wide field of application such as non-destructive evaluations~\cite{JILES1988311} 
and medical imaging based on nuclear spin magnetic resonance (NMR).
In particular, qubit-based sensors, called quantum sensors, have attracted considerable attention 
for their ultra-high sensitivity beyond what is achievable with conventional magnetic sensors
such as semiconductor-based Hall sensors.~\cite{Degen2008}
The quantum sensors can achieve highly sensitive measurements, 
including magnetometry,~\cite{Taylor2008, doi:10.1146/annurev-physchem-040513-103659, 0034-4885-77-5-056503}
electrometry,~\cite{Dolde2011, PhysRevA.95.053417, Ariyaratne2018} 
and thermometry.~\cite{Kucsko2013, Toyli8417, PhysRevB.91.155404}
A quantum sensor can also detect a very small mechanical oscillation.~\cite{Kolkowitz1603}

Among the quantum sensors developed so far, a nitrogen-vacancy (NV) quantum sensor,
where the electron-spin state of an NV center in diamond is used as a spin qubit,
has high spatial resolution~\cite{Degen2008} in addition to high sensitivity.
It can therefore detect magnetism~\cite{Balasubramanian2008, Rondin2013, Gross2017, Yane1603137} 
and nuclear spins~\cite{Mamin557, Staudacher561, doi:10.1021/nl402286v, Lovchinsky503, Aslam67}
in a nanometer scale.
For such applications, the amplitude of an alternating-current (AC) magnetic field is generally measured 
using AC magnetometry techniques with multiple-pulse dynamical decoupling sequences.
In these techniques, the maximum phase accumulation of the NV spin due to an AC magnetic field can be obtained
when the period of $\pi$ pulses in the decoupling sequences matches the half time period of the AC field
and the relative phase difference between the AC field and the decoupling sequences is zero.    
This phase-accumulation condition can enhance the signal from a particular-frequency field 
and suppress the influence of noise caused by unwanted-frequency fields,
which can be regarded as a function of a lock-in amplifier in the quantum regime.~\cite{Kotler2011}
Consequently, the NV quantum sensor can detect an AC magnetic field with a very small amplitude.~\cite{Taylor2008, 
doi:10.1146/annurev-physchem-040513-103659, Balasubramanian2009} 
Furthermore, such amplitude measurement is also utilized for vector-field sensing and imaging.~\cite{1367-2630-13-4-045021} 
By contrast, phase accumulation of the NV spin is not obtained when the relative phase difference is $\pi/2$.
This non-phase-accumulation condition is therefore considered unfavorable for measuring the amplitude of an AC magnetic field.
In this study, however, we found that under the non-phase-accumulation condition,
the NV quantum sensor can measure the phase shift of an AC magnetic field from its initial phase with high sensitivity.
Therefore, the amplitude and the phase shift of an AC magnetic field can be sensitively measured 
by performing NV quantum sensing under phase-accumulation and non-phase accumulation conditions, respectively.

\section*{Results}
\subsection*{AC magnetometry}
\begin{figure}[htbp]
	\begin{center}
		\includegraphics[keepaspectratio, width = 12cm]{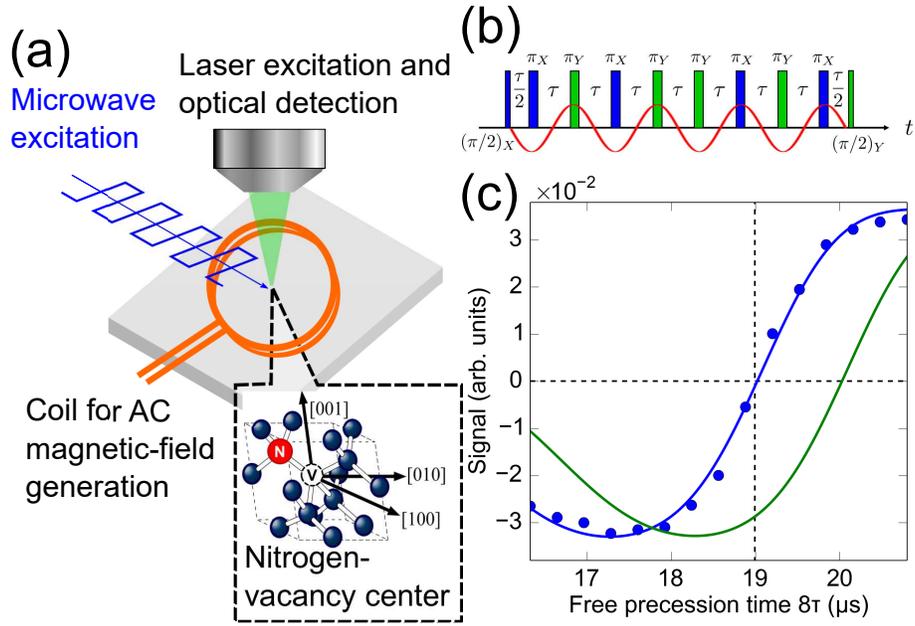}		
	\end{center}
	\caption{(a) Schematic of the experimental setup for the AC magnetometry
				based on a quantum sensor with multiple-pulse decoupling sequences.
				As a quantum sensor, we used an ensemble of the electron-spin states
				of NV centers in an isotopically purified diamond film.
				Details are described in Methods.
				(b) An XY8-1 decoupling sequence under the non-phase-accumulation condition. 
				The relative phase difference between the decoupling sequences
				and the AC magnetic field (red solid line) is $\pi/2$ shifted.
				The blue and green bars show different microwave phases, 
				which are in quadrature to each other.
				The number of $\pi$ pulses is $N = 8$, 
				and the first $\pi/2$ pulse is in quadrature 
				with another $\pi/2$ pulse subsequent to the $\pi$-pulse train.
				(c) AC magnetometry result using the XY8-1 sequence. 
				Details are explained in the main text.
				The horizontal dashed line indicates that 
				the NV quantum sensor acquires no phase accumulation of the field, 
				and the vertical dashed line indicates the free precession time $N\tau$ 
				satisfying $\tau + \tau_{\pi} = \left ( 2f_{ac} \right )^{-1}$,
				where $\tau_{\pi}$ is the $\pi$ pulse width.
				In this study, the pulse width was set at $\tau_{\pi} = 124~\mathrm{ns}$. 
				}
	\label{fig:XY8}	
\end{figure}

We realized a quantum sensor
by using an ensemble of the electron-spin states of NV centers in an isotopically purified diamond film.
Figure~\ref{fig:XY8}(a) shows a schematic of our experimental setup for the AC magnetometry. 
An AC magnetic field of 
$B(t)=B_{ac} \cos \left ( 2\pi f_{ac} t + \phi_{ac} \right )$ was generated by the coil,
where $B_{ac}$ is the amplitude, $f_{ac}$ is the frequency, 
and $\phi_{ac}$ is the initial phase at the start of the decoupling sequences.
The relative phase difference between the field and the decoupling sequences can be controlled
by tuning the initial phase $\phi_{ac}$. 
Here, we set $f_{ac} = 200~\mathrm{kHz}$ and $\phi_{ac} = \pi/2$. 
The AC field was applied to the NV ensemble and 
synchronized to the decoupling sequences, as shown in Fig.~\ref{fig:XY8}(b).

Figure~\ref{fig:XY8}(c) shows 
the AC magnetometry result obtained using an XY8-1 sequence.~\cite{GULLION1990479}
The XY-series sequences are classified as a Carr--Purcell (CP)-type sequence.~\cite{PhysRev.94.630}
Therefore, the signal of AC magnetometry 
obtained by using this type of decoupling sequences with even $N$ pulses is proportional to 
$\sin \Phi$, where $\Phi$ is given by~\cite{1807.00946}
\begin{eqnarray}
	\Phi &=&
	\gamma_{e} B_{ac} 
	\cos \left [
	\pi f_{ac} N \tau (1 + \alpha) + \phi_{ac}  
	\right ] \nonumber \\
	&~& 
	\times
	N \tau (1 + \alpha)
	\left \{
	1 - \frac { \cos (\alpha \pi f_{ac} \tau ) }
	{\cos [ \pi f_{ac} \tau (1 + \alpha)  ] } 
	\right \}
	\frac{
	\sin  \left [ \pi f_{ac} N \tau (1 + \alpha)   \right ]
	}
	{\pi f_{ac} N \tau (1 + \alpha) }.
	\label{eq:PhaseAccCP}
\end{eqnarray}
Here, $\gamma_e$ is the gyromagnetic ratio of the electron-spin states of NV centers,
$N$ is the number of $\pi$ pulses, 
$\tau$ is the interval between the pulses,
and $\alpha$ is the ratio of the $\pi$-pulse width $\tau_{\pi}$ to the interval $\tau$, 
i.e., $\tau_{\pi} = \alpha \tau$.
In this study, $\tau_{\pi} = 124~\mathrm{ns}$.
Fitting the experimental results according to $\sin \Phi$ [Eq.~\eqref{eq:PhaseAccCP}] 
yielded $B_{ac} = 0.74 \pm 0.02~\mathrm{\mu T}$ 
and $\phi_{ac} = 1.55 \pm 0.02 \approx \pi/2$.
This fitting result is indicated by the blue solid line in Fig.~\ref{fig:XY8}(c).
Note that in an ideal case with $\tau_{\pi} = 0$ (i.e., $\alpha = 0$), 
no phase accumulation should be observed at the free precession time $N\tau = 20~\mathrm{\mu s}$ 
where $N = 8$ and $\tau = (2f_{ac})^{-1}$. This is indicated by the green solid line in Fig.~\ref{fig:XY8}(c).
In the experiment, however, no phase accumulation was observed 
at the time $\approx 19~\mathrm{\mu s}$ because of the finite $\pi$-pulse width; 
therefore, $\tau$ should be modified as $\tau = [2 f_{ac} (1+\alpha)]^{-1}$.
The details of the finite-width-pulse effect have been described by our previous work.~\cite{1807.00946}

\subsection*{Signal deviation due to phase shift}
The magnetometry-signal deviation $\Delta S$ for a very small phase shift $\Delta \phi_{ac}$ from $\phi_{ac}$ is 
obtained in terms of the derivative of a mangetometry signal $\sin \Phi$ with respect to $\phi_{ac}$:
\begin{equation}
\label{eq:DeltaSignal}
\centering
\Delta S \propto \Delta \phi_{ac} \frac{\partial \Phi }{\partial \phi_{ac} } \cos\Phi.
\end{equation}
The proportionality factor in Eq.~\eqref{eq:DeltaSignal} is explained in Methods.
An important feature of this equation is that $\Delta S$ is proportional to $\cos{\Phi}$. 
This implies that the maximum $\Delta S$ at $\tau = [2 f_{ac} (1+\alpha)]^{-1}$ 
can be obtained under the non-phase-accumulation condition,
where the relative phase difference between the AC field and the decoupling sequences is $\pi/2$, 
i.e., $\phi_{ac} = \pi/2$ in this study. 

\begin{figure}[htbp]
	\begin{center}
		\includegraphics[keepaspectratio, width = 8.5cm]{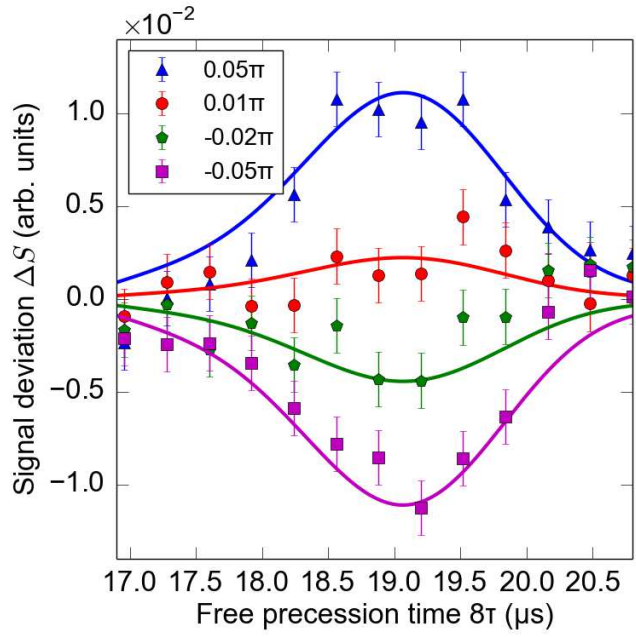}		
	\end{center}
	\caption{Deviation of AC magnetometry signal obtained using the XY8-1 sequence
				as a function of the free precession time $N \tau$ with $N = 8$.
				We measured the deviation
				from the magnetometry signal at $\phi_{ac} = \pi/2$
				by using several phase shifts $\Delta \phi_{ac}$:
				$\Delta \phi_{ac} =0.05\pi$ (blue triangles), $0.01\pi$ (red circles), 
				$-0.02\pi$ (green pentagons) and $-0.05\pi$ (purple squares).
				Error bars in the figure indicate the standard deviation of the photon shot noise.
				}
	\label{fig:PhaseDev}	
\end{figure}

Figure~\ref{fig:PhaseDev} shows $\Delta S$ for the AC magnetometry obtained by using the XY8-1 sequence
with various phase shifts $\Delta \phi_{ac}$ for the initial phase $\phi_{ac} = \pi/2$:
$\Delta \phi_{ac}$ = $\pm 0.05\pi$, $-0.02\pi$, and $0.01\pi$.
In this experiment,
the same parameters of the AC field as those for the previous experiment of Fig.~\ref{fig:XY8} were adopted.
As expected from Eq.~\eqref{eq:DeltaSignal}, 
the maximum $\Delta S$ for each $\Delta \phi_{ac}$ was obtained around $19~\mathrm{\mu s}$, 
which is the non-phase-accumulation condition (see Figs.~\ref{fig:XY8}(b) and (c)).
We also confirmed that the experimental results can be reproduced by Eq.\eqref{eq:DeltaSignal} 
with the AC-field and the decoupling-sequence parameters and $\Delta \phi_{ac}$, 
as indicated by the solid lines in Fig.~\ref{fig:PhaseDev}.

\begin{figure}[htbp]
	\begin{center}
		\includegraphics[keepaspectratio, width = 8.5cm]{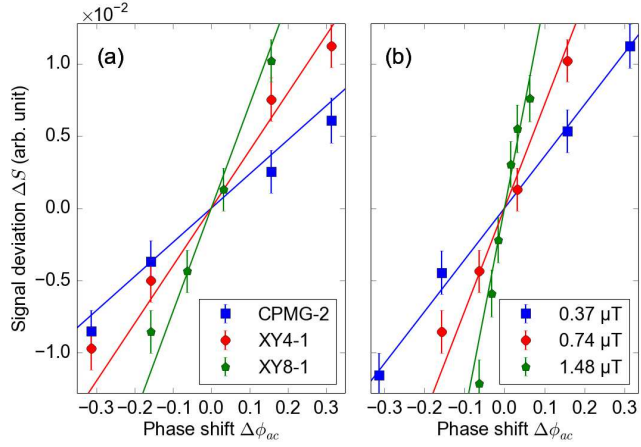}		
	\end{center}
	\caption{Pulse-number and field-amplitude dependences of the magnetometry-signal deviations
				as a function of the phase shift from the initial AC-field phase $\phi_{ac}$. 
				Error bars in the figure indicate the standard deviation of the photon shot noise.
				Here, the free precession time $N\tau$ was fixed,
				where $\tau(1+\alpha) = \left ( 2f_{ac} \right ) ^{-1}$.
				(a) The dependence of the magnetometry signal deviation on the number of $\pi$ pulses.
				We obtained the AC magnetometry data using
				CPMG-2 ($N = 2$;  blue squares), 
				XY4-1 ($N = 4$; red circles), and XY8-1 ($N = 8$; green pentagons).
				Each colored solid line represents each experimental result
				according to Eq.~\eqref{eq:DerPhi}.
				(b) The dependence of the magnetometry signal deviation on the field amplitude. 
				Here, we performed AC magnetometry measurements using the XY8-1 sequence 
				at a fixed free precession time $8\tau \approx 19~\mathrm{\mu s}$ 
				and for $B_{ac} = 0.37~\mathrm{\mu T}$ (blue squares),
				$0.74~\mathrm{\mu T}$ (red circles), and $1.48~\mathrm{\mu T}$ (green pentagons).
				They are consistent with our theoretical plots [Eq.~\eqref{eq:DerPhi}]
				indicated by each colored solid line.
				}
	\label{fig:PiNum_Mag}	
\end{figure}

If $\phi_{ac} = \pi/2$ and $\tau(1+\alpha) = \left ( 2f_{ac} \right ) ^{-1}$
(i.e., the non-phase-accumulation condition), 
$\Delta S/\Delta \phi_{ac}$ for approximately $\Phi = 0$ is given by
\begin{eqnarray}
\centering
\frac{\Delta S}{\Delta \phi_{ac}} \propto  
\frac{\partial \Phi}{\partial \phi_{ac}} 
=
(-1)^{N+1}\frac{2}{\pi} \gamma_{e} B_{ac} N \tau (1+\alpha).
\label{eq:DerPhi}
\end{eqnarray}
Therefore, $\Delta S$ is expected to be proportional to the number of pulses $N$ 
and the AC field amplitude $B_{ac}$.
To examine the dependence of $\Delta S$ on $N$, 
we performed AC magnetometry using a Carr--Purcell--Meiboom--Gill~\cite{doi:10.1063/1.1716296}
sequence with $N=2$ (CPMG-2),
a XY4-1 sequence ($N = 4$), and the XY8-1 sequence ($N = 8$)
at a fixed free precession time $N\tau$, 
where $\tau(1+\alpha) = \left ( 2f_{ac} \right ) ^{-1} = 2.5~\mathrm{\mu s}$.
Here, we fixed the AC-field amplitude at $B_{ac} = 0.74~\mathrm{\mu T}$. 
Figure~\ref{fig:PiNum_Mag}(a) shows $\Delta S$ 
as a function of phase shift $\Delta \phi_{ac}$ for each decoupling sequence.
The solid lines in Fig.~\ref{fig:PiNum_Mag}(a) were obtained by Eq.~\eqref{eq:DerPhi}
with the AC-field and the decoupling-sequence parameters.
We observed that the slopes of the $\Delta S$--$\Delta \phi_{ac}$ relations increased
with increasing $N$; this agrees with Eq.~\eqref{eq:DerPhi}.
We also examined the dependence of $\Delta S$ on $B_{ac}$ for the XY8-1 sequence
at a fixed free precession time $8 \tau \approx 19~\mathrm{\mu s}$ (non-phase-accumulation condition).
Figure~\ref{fig:PiNum_Mag}(b) shows that 
the slopes of the $\Delta S$--$\Delta \phi_{ac}$ relations increased with increasing $B_{ac}$;
this is consistent with Eq.~\eqref{eq:DerPhi}.

\section*{Discussion}
\subsection*{Signal-to-noise ratio for phase-shift measurement}
As mentioned previously, the maximum phase accumulation can be acquired 
when $\phi_{ac} = 0$ and  $\tau (1 + \alpha ) = \left ( 2f_{ac} \right ) ^{-1}$.
By contrast, our results indicate that 
under the non-phase-accumulation condition, where $\phi_{ac} = \pi/2$, 
the NV quantum sensor can detect a very small phase shift of the AC magnetic field.
Assuming that a detection error is caused by the shot noise of photons detected from the NV ensemble, 
the signal-to-noise ratio (SNR) for the $\Delta \phi_{ac}$ measurement is given by
\begin{equation}
\label{eq:SNR}
\mathrm{SNR} \sim \sqrt{N_{m} N_{nv}} C 
\exp \left \{ - \left [ \frac{N\tau}{T_2(N)} \right ]^{p} \right \}
\left | \frac{\partial \Phi }{ \partial \phi_{ac} } \right | |\Delta \phi_{ac}|,
\end{equation}
where $N_{m}$ is the number of measurements, $N_{nv}$ is the number of NV centers in an optical detection volume,
and $T_2(N)$ is the coherence time of the NV ensemble with the N-pulse decoupling sequences.
$ C = \left [ 1 + 2(r_{0} + r_{1})/(r_{0} - r_{1} ) ^2 \right ] ^{-1/2}$, where $r_{0}$ and $r_{1}$
are photon counts of the bright and the dark states of NV centers, respectively.
The derivation of Eq.~\eqref{eq:SNR} is described in Methods.
In our experimental setup, $C \approx 0.03$ and $N_{nv} \approx 60$.
If we measure $\Delta \phi_{ac}$ of an AC magnetic field in $N\tau \sim T_2(N)$, 
we can define the detectable minimum phase shift $\Delta \phi_{ac}^{\mathrm{min}}$
according to Eqs.~\eqref{eq:DerPhi} and \eqref{eq:SNR} with $\mathrm{SNR} = 1$, as follows:
\begin{eqnarray}
\centering
\left | \Delta \phi_{ac} ^{\mathrm{min}} \right |
= \frac{\eta_{\phi}}{\sqrt{T_t}} 
\sim
\frac{1}{\sqrt{T_t}} \frac{1}{\sqrt{N_{nv}}}
\frac{\pi \exp(1)}{ 2 C \gamma_e B_{ac} \sqrt{T_2(N)} }, 
\label{eq:PhaseSensitivity}
\end{eqnarray}
where the total measurement time $T_{t} = N_{m} N\tau (1+ \alpha) \sim N_{m} T_2(N)$ if $\alpha$ is neglected.
Defining the phase sensitivity $\eta_{\phi}$ as a formula similar to that of the magnetic-field sensitivity in the quantum sensing,~\cite{Taylor2008} 
we can estimate $\eta_{\phi} \approx 3 \times 10^{-3} /\sqrt{\mathrm{Hz}}$
with $B_{ac} = 1~\mathrm{\mu T}$ and $T_2(N=256) \approx 1~\mathrm{m s}$ (See the subsection "Measurements of coherence time" in Methods).

\subsection*{Upper limit of the $\Delta \phi_{ac}$ measurement}
Note that Eqs.~\eqref{eq:DeltaSignal}--\eqref{eq:PhaseSensitivity} about the phase shift  of the AC field 
are valid only for the condition that $\Delta S$ is (linearly) proportional to $\Delta \phi_{ac}$. 
However, as $\Delta \phi_{ac}$ increases, 
the linear relationship between $\Delta S$ and $\Delta \phi_{ac}$ is no longer maintained. 
Therefore, Eqs.~\eqref{eq:DeltaSignal}--\eqref{eq:PhaseSensitivity} are applicable for $\left | \Delta \phi_{ac} \right | \ll 1$.

\begin{figure}[htbp]
	\begin{center}
		\includegraphics[keepaspectratio, width = 8.5cm]{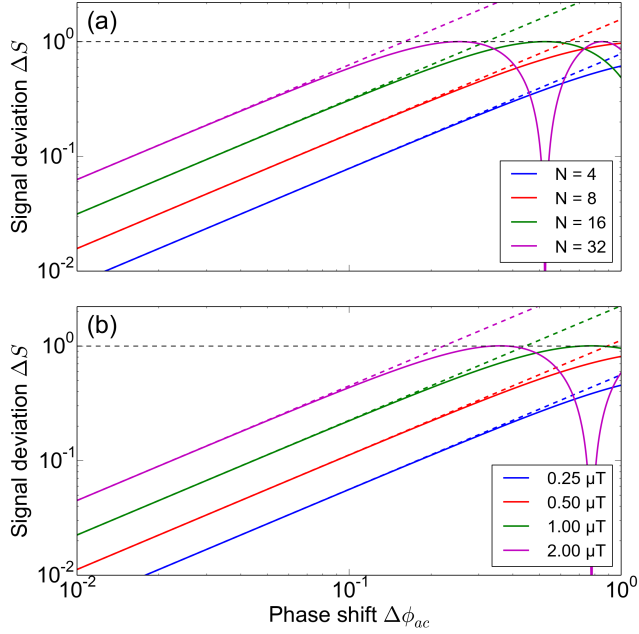}		
	\end{center}
	\caption{	Absolute values of the signal deviation $\left | \Delta S \right |$
			    	obtained from $| \sin{\Phi(\phi_{ac} + \Delta \phi_{ac})} 
				- \sin{\Phi(\phi_{ac})} |$ (colored solid lines) 
				as well as those obtained from Eq.~\eqref{eq:DerPhi} (colored dashed lines) .
				They are plotted as a function of $\Delta \phi_{ac}$ for various $N$ and $B_{ac}$
				in (a) and (b), respectively. 
				Each limit of the phase-shift measurement 
				$\left | \Delta \phi_{ac}^{L} \right |$ is indicated by the phase-shift value of the intersection point 
				between the black and each colored-dashed line.
				(a) Dependence of the magnetometry signal deviation on $N$. Here, we fixed the field amplitude 
				at $B_{ac} = 0.7~\mathrm{\mu T}$.
				(b) Dependence of the magnetometry signal deviation on the amplitude $B_{ac}$ if $N=8$.
				}
	\label{fig:PhaseShiftRange}	
\end{figure}

To evaluate a range of $\Delta \phi_{ac}$ 
for which Eqs.~\eqref{eq:DeltaSignal}--\eqref{eq:PhaseSensitivity} are applicable for the estimation of $\Delta \phi_{ac}$,
in Figs.~\ref{fig:PhaseShiftRange}(a) and \ref{fig:PhaseShiftRange}(b), respectively,
we plotted the absolute values of the signal deviation $\left | \Delta S \right |$
obtained from $\left | \sin{\Phi(\phi_{ac} + \Delta \phi_{ac})} - \sin{\Phi(\phi_{ac})} \right |$ (colored solid lines) 
along with those obtained from Eq. (3) (colored dashed lines) 
as a function of $\Delta \phi_{ac}$ for various $N$ and $B_{ac}$ values.
Here, the intersection points $\Delta \phi_{ac}^{L}$ 
between the colored dashed line and $\left | \Delta S \right | = 1$ (black dashed lines) are given by
\begin{equation}
\label{eq:boundary}
\centering
\left | \Delta \phi_{ac}^{L} \right |
=
\frac{\pi/2}{\gamma_e B_{ac} N \tau (1 + \alpha)}.
\end{equation}
Note that $\left | \Delta S \right | = 1$ corresponds to the maximum $\left | \Delta S \right |$ 
obtained from $| \sin{\Phi(\phi_{ac} + \Delta \phi_{ac})} - \sin{\Phi(\phi_{ac})} |$
when the proportionality factor is assumed to be unity;
thus, $\Delta \phi_{ac}^{L}$ can be used as a measure of the upper limit of $\Delta \phi_{ac}$
at which Eqs.~\eqref{eq:DeltaSignal}--\eqref{eq:PhaseSensitivity} are applicable.
As seen in Fig.~\ref{fig:PhaseShiftRange}, for 
$\left | \Delta \phi_{ac} \right | \ll \left | \Delta \phi_{ac}^{L} \right |$,
$\left | \Delta S \right |$ obtained from 
$| \sin{\Phi(\phi_{ac} + \Delta \phi_{ac})} - \sin{\Phi(\phi_{ac})} |$
is practically proportional to $\Delta \phi_{ac}$, 
implying that Eq.~\eqref{eq:DerPhi} is applicable for estimating $\Delta \phi_{ac}$.
As discussed previously, $\left | \Delta S \right |$ increases with increasing $N$ and $B_{ac}$ (Fig.~\ref{fig:PiNum_Mag}).
This means that the measurement sensitivity of $\Delta \phi_{ac}$ can be improved by increasing $N$ and $B_{ac}$.
However, Fig.~\ref{fig:PhaseShiftRange} shows $\left |  \Delta \phi_{ac}^{L} \right |$ reduces with increasing $N$ and $B_{ac}$,
indicating that the upper limit $\Delta \phi_{ac}$
at which Eqs.~\eqref{eq:DeltaSignal}--\eqref{eq:PhaseSensitivity} are applicable
reduces with increasing $N$ and $B_{ac}$.
Therefore, in the estimation of $\Delta \phi_{ac}$ by the AC magnetometry using multiple-pulse decoupling sequences,
appropriate $N$ and $B_{ac}$ that depend on $\Delta \phi_{ac}$ exist.

\subsection*{Future prospect}
In this study, we found that NV quantum sensing based on multiple-pulse decoupling sequences
can detect a very small phase shift $\Delta \phi_{ac}$ of an AC magnetic field 
under the non-phase-accumulation measurement condition.
The sensitivity of $\Delta \phi_{ac}$ measurement depends on the number of $\pi$ pulses ($N$)
and the amplitude of an AC magnetic field ($B_{ac}$).
The results of the experiments conducted in this study reveal a measurable $\Delta \phi_{ac}$ of approximately $10^{-3}$.
Here, the measurable $\Delta \phi_{ac}$ was found to depend 
on the number of NV centers in a detection volume $N_{nv}$ and their coherence time $T_2(N)$.
Recently, Wolf et al. have reported $T_2(N=1) = 100~\mathrm{\mu s}$ for $N_{nv} \approx 10^{11}$,
which can realize subpicotesla magnetometry.~\cite{PhysRevX.5.041001}
Substituting these values to Eq.~\eqref{eq:PhaseSensitivity},
we can expect that, for a NV quantum sensing, 
the measurable $\Delta \phi_{ac}$ reaches to the order of $10^{-7}$ for $B_{ac} = 1~\mathrm{\mu T}$.
However, in addition to the optical detection error, 
the measurable $\Delta \phi_{ac}$ should be affected by a variation in the AC-field amplitude
and an error in the quantum control of NV-spin states. 
Therefore, to realize an accurate $\Delta \phi_{ac}$ measure,
we must further investigate what causes its uncertainty.
In this study, we used the conventional multiple-pulse decoupling sequences 
developed for the phase accumulation of spins so far.
However, other types of multiple-pulse decoupling sequences may be appropriate for measuring $\Delta \phi_{ac}$,
because the phase measurement is performed under the non-phase-accumulation condition. 
Therefore, further developments on multiple-pulse decoupling sequences are also needed 
for realizing an accurate measurement of $\Delta \phi_{ac}$.

The proposed $\Delta \phi_{ac}$ measurement technique can be applied 
for measuring a complex AC magnetic field in the micrometer and nanometer scales.
Since the dynamical decoupling sequences utilizing multiple-pulses decoupling sequences are commonly used 
for quantum information processing,~\cite{PhysRevLett.82.2417, PhysRevB.77.174509, PhysRevA.79.062324, Bylander2011, Kawakami11738}
they are applicable for quantum sensing irrespective of the type of qubtis,
which makes the proposed technique applicable for other qubits 
such as superconducting circuits.~\cite{cond-mat/0411174, 0034-4885-80-10-106001, GU20171}
Recently, qubits have been utilized for detection of non-classical fields such as squeezed states.~\cite{PhysRevLett.119.023602,PhysRevLett.120.040505,PhysRevX.7.041011,Clark2016}
Our findings will provide a guide for realizing a sensitive measurement of quantum noise~\cite{RevModPhys.82.1155}
caused by a phase shift of such non-classical fields
and will contribute to advances in quantum measurements based on qubits 
through the sensitive measurement of phase shifts of a complex AC magnetic field.

\section*{Methods}
\subsection*{Sample information}
We used an ensemble of NV centers 
($\approx 3 \times 10^{14}~\mathrm{cm^{-3}}$)
in an isotopically purified diamond film.
The isotopically controlled $^{12}$C diamond film was prepared 
by microwave-plasma-assisted chemical vapor deposition 
from isotopically controlled $\mathrm{ ^{12}C H_{4}}$  ($ > 99.999~\%$ for $\mathrm{ ^{12}C} $)
and $\mathrm{H_2}$ mixed gas.
We used a reactant gas with a nitrogen to carbon ratio $\mathrm{N}/\mathrm{C}=1.75~\%$ during the growth
to produce the ensemble of NV centers.
The details of sample preparation have been reported elsewhere.~\cite{7466817, 1807.00946}

\subsection*{Experimental procedure and setup}
Before applying the dynamical decoupling sequences,
we initialized the NV-spin states by using a $532~\mathrm{nm}$ laser pulse of $5~\mathrm{\mu s}$.
Microwave pulses were applied to the NV quantum sensor for the multiple-pulse decoupling sequences.
After applying the decoupling sequences, we obtained the AC magnetometry signals 
by detecting of photons from the NV quantum sensor excited by a laser pulse ($500~\mathrm{ns}$). 
We canceled the common-mode noise of our measurements based on dynamical decoupling sequences
by using a $3\pi/2$ pulse instead of a $-\pi/2$ pulse.\cite{Bar-Gill2013}

In this study, the AC magnetometry was performed 
by using a home-built laser scanning microscope with a 0.85 numerical aperture objective (Olympus LCPLFLN100x LCD).
The optical initialization and detection of the NV-spin states were performed
by a $532~\mathrm{nm}$ continuous wave laser  (SOC Showa Optoronics J150GS-1G-12-23-12) coupled 
with two acousto-optic modulators (Optoelectronic MQ180-G13-FIO) for optical pulse operation
with a high extinction ratio ($>~80~\mathrm{dB}$).
Photons from the NV quantum sensor were detected by a single photon counter (Excelitas Technologies SPCM-AQRH-15-FC).
Our microwave setup consists of a microwave source (QuickSyn Synthesizers FSW-0020)
and a quadrature hybrid coupler (Marki QH-0R714 Quadrature Hybrid Coupler)
with each output connected to a microwave switch (Mini-Circuits ZYSWA-2-500DR) 
for generating either in-phase or quadrature-phase microwave pulses.
After each switch, both output paths were combined
and then amplified by Mini-Circuits ZHL-16W-43+, followed by a microwave antenna.~\cite{doi:10.1063/1.4952418}
A static magnetic field of $\approx 4~\mathrm{mT}$ was applied with 
its direction parallel to a $\left \langle 111 \right \rangle$ axis of the diamond.
An AC magnetic field was applied to the NV ensemble by a home-made coil 
connected to an arbitrary function generator (Tektronix AFG3252),
which was used to control the frequency, initial phase, and amplitude of the AC magnetic field.
A TTL pulse generator (SpinCore PulseBlaster ESR-PRO-500) controlled
timing of the pulse sequences and synchronization between the AC magnetic field and the sequences.

\subsection*{Measurements of coherence time}
\begin{figure}[htbp]
	\begin{center}
		\includegraphics[keepaspectratio, width = 8.5cm]{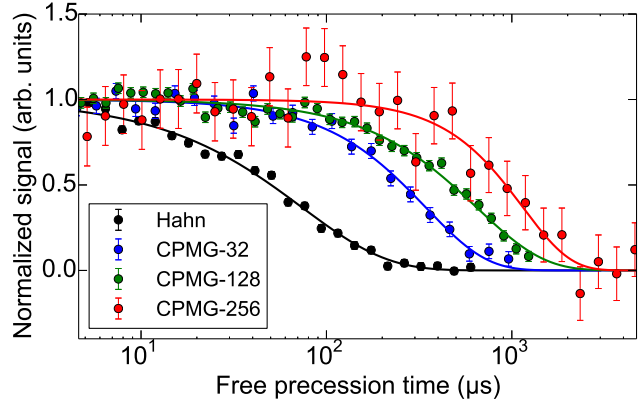}		
	\end{center}
	\caption{Measurement results using the Hahn-echo (black), 
				CPMG-32 (blue), CPMG-128 (green), and CPMG-256 (red) sequences.
				Error bars are obtained from the standard deviation of the photon shot noise.
				Each colored solid line corresponds 
				to each fitting result obtained with Eq.~\eqref{eq:Coherence}, 
				where the parameters are shown in Tab.~\ref{tab:Fitting}. 
				}
	\label{fig:Coherence}	
\end{figure}

The coherence time of the NV ensemble was measured
to estimate the phase sensitivity given by Eq.~\eqref{eq:PhaseSensitivity} in the main text.
As shown in Fig.~\ref{fig:Coherence}, we measured coherence time by using the Hahn-echo, 
CPMG-32, CPMG-128, and CPMG-256 sequences. 
In these measurements, 
the first and the last $\pi/2$ pulses in each sequence were in-phase 
and the AC field was not applied.
We fit the results with
\begin{equation}
\label{eq:Coherence}
\mathrm{Signal}
\propto
\exp{
\left [
- \left ( \frac{N\tau}{T_2(N)} \right ) ^ p
\right ]
},
\end{equation}
where 
 $T_2(N)$ is the coherence time of the NV ensemble in decoupling sequences with N pulses.
The fitting parameters substituted in Eq.~\eqref{eq:Coherence} 
for reproducing the results are described in Tab.~\ref{tab:Fitting}.
According to the experimental results,
we substituted $T_2(N = 256) \approx 1~\mathrm{m s}$ in Eq.~\eqref{eq:PhaseSensitivity} for estimating the phase sensitivity.

\subsection*{Proportionality factor in Eqs.~\eqref{eq:DeltaSignal} and \eqref{eq:DerPhi}}
We used the common-mode-rejection method~\cite{Bar-Gill2013} to obtain the results;
thus, the below gives the signal of AC magnetometry obtained by applying multiple-pulse sequences, 
where the first $\pi/2$ pulse is quadrature-phase to the last $\pi/2$ pulse:~\cite{1807.00946}
\begin{equation}
\label{eq:Signal_q}
S = (-1)^{n_y +1}\frac{1-r}{1+r} 
\exp{
\left [
- \left ( \frac{N\tau}{T_2(N)} \right ) ^ p
\right ]
}
\sin{\Phi}.
\end{equation} 
Here, $r$ is the ratio of photon counts from the dark to the bright states of the NV center,
$n_y$ is the number of $\pi_Y$ pulses, 
and $T_2(N)$ is the coherence time of $N$-pulse decoupling sequences.
$\Phi$ is the phase accumulation due to an AC magnetic field
and is given by Eq.~\eqref{eq:PhaseAccCP} in the main text.
Therefore, the proportionality factor in Eqs.~\eqref{eq:DeltaSignal} and \eqref{eq:DerPhi} is given by
\begin{equation}
\label{eq:Coef}
(-1)^{n_y +1}\frac{1-r}{1+r} 
\exp{
\left [
- \left ( \frac{N\tau}{T_2(N)} \right ) ^ p
\right ]
}.
\end{equation}
To obtain the parameters in Eq.~\eqref{eq:Coef}, 
we measured the coherence time by using the XY8-1 sequence,
where both the $\pi/2$ pulses were in-phase and the AC field was not applied.
From the measurements, 
$T_2(N=8) = 140\pm10~\mathrm{\mu s}$, $p = 0.97 \pm 0.09$, and $r = 0.917 \pm 0.001$. 
These obtained parameters were used 
to fit the experimental data with Eq.~\eqref{eq:Signal_q}, as shown in Fig.~\ref{fig:XY8}(c),
and to plot theoretical results given 
by Eqs.~\eqref{eq:PhaseAccCP}--\eqref{eq:DerPhi}. 
The obtained parameters of the CPMG-2, XY4-1, and XY8-1 sequences are shown 
in Tab~\ref{tab:Parameters}.

\subsection*{Signal-to-noise ratio for phase-shift measurement}
To derive the SNR for the phase-shift measurement, 
we calculated the expectation value and the variance of an AC magnetometry signal 
by using a measurement operator and a density matrix of the NV-spin states.
Our derivation is based on the theoretical procedure reported by Meriles et al.\cite{doi:10.1063/1.3483676}

Assuming that NV-spin states are represented as a two-level system,
the measurement operator related to the optical detection is defined by 
\begin{equation}
\label{eq:MeasOp}
M = a \left | 0 \rangle \langle 0 \right |
+ b \left | 1 \rangle \langle 1 \right | 
=
\frac{ a + b }{2} I
+
\frac{ a - b }{2} \sigma_z,
\end{equation}
where $\left | 0 \right \rangle$ ($\left | 1 \right \rangle$) corresponds to the bright (dark) state of NV centers.
$\sigma_z$ is the Pauli matrix and $I$ is the identity operator.
Here, $a$ and $b$ are two independent and stochastic variables of the bright and dark states
associated with collecting photon counts measured by optical readout pulses.
These are characterized by Poisson distributions, i.e., 
$\left \langle a \right \rangle = r_0$ and $\left \langle b \right \rangle = r_1$.
In addition, the ratio $r = r_1 /r_0$, 
as shown in Eqs.~\eqref{eq:Signal_q} and \eqref{eq:Coef}.
Because of the optical initialization,
the initial density matrix of the NV-spin states is given by
\begin{equation}
\label{eq:InitDensityOp}
\rho_i = \left | 0 \rangle  \langle 0 \right |
= \frac{1}{2} \left ( 1 + \sigma_z  \right ).
\end{equation}
By applying the first $(\pi/2)_X$ pulse,
the density matrix is given by
\begin{equation}
\label{eq:DensityOp0}
\rho(t=0) =  \frac{1}{2} \left ( 1 - \sigma_y \right ).
\end{equation} 
In this study, an AC field was applied to the NV centers subsequent to the $(\pi/2)_X$ pulse
and synchronized to the decoupling sequences with multiple $\pi$ pulses, 
as shown in Fig.~\ref{fig:XY8}(b).
After applying the decoupling sequences with the $\pi$ pulses, which is synchronous to the AC magnetic field,
the density matrix is given by
\begin{equation}
\label{eq:DensityOpT}
\rho(T)= \frac{1}{2} 
\left \{ 
1 - (-1)^{n_x} \sigma_y \cos \Theta(T) +
(-1)^{n_y} \sigma_z \sin \Theta(T) 
\right \} ,
\end{equation}
where $n_x$ and $n_y$ are the number of $\pi_X$ and $\pi_Y$ pulses, respectively.
Here, $T = N\tau(1 + \alpha) $ and $n_x + n_y = N$.
In Eq.~\eqref{eq:DensityOpT}, 
$\Theta$ involves the phase accumulation due to the AC field
and magnetic-field noise from the local environment around the NV centers.
The former case is given by $\Phi$ [Eq.~\eqref{eq:PhaseAccCP}], 
and the latter one causes decoherence of the NV-spin states.
After applying the $(\pi/2)_Y$ pulse subsequent to the $\pi$-pulse train,
the final density matrix before the optical detection is given by
\begin{equation}
\label{eq:FinalDensityOp}
\rho_f= \frac{1}{2} 
\left \{ 
1 - (-1)^{n_x} \sigma_y \cos \Theta(T) -
(-1)^{n_y} \sigma_z \sin \Theta(T) 
\right \}.
\end{equation}
Therefore, the expected measurement operator value is represented as follows:
\begin{eqnarray}
\left  \langle M \right \rangle 
&= &
\mathrm{Tr}  \left \{ M \rho_f \right \} \nonumber \\
&=&\frac{r_0 + r_1 }{2}
+(-1)^{n_y+1} \frac{ r_0 - r_1 }{2}
\exp{\left (-D \right )}\sin{\Phi(T)}, 
\end{eqnarray}
where the decoherence factor $D = [N \tau/T_2(N)]^p$. 
$\Phi$ is given by Eq.~\eqref{eq:PhaseAccCP}.
According to the derivative of $\left \langle M \right \rangle$
with respect to an AC-field phase $\phi_{ac}$, 
the signal deviation $\Delta S$ for the phase shift $\Delta \phi_{ac}$ is represented by 
\begin{eqnarray}
\Delta S &=& (-1)^{n_y+1} \frac{ r_0 - r_1 }{2} \exp{\left (-D \right )}
\Delta \phi_{ac} \left ( \frac{\partial}{\partial \phi_{ac}}\sin{\Phi(T)}  \right ) \nonumber \\
&=& (-1)^{n_y+1} \frac{ r_0 - r_1 }{2} \exp{\left (-D \right )}
\Delta \phi_{ac} \cos{ \Phi(T) } \left ( \frac{\partial \Phi(T)}{\partial \phi_{ac}} \right ).
\label{eq:DerSignal}
\end{eqnarray} 
On the other hand, the variance of the measurement operator is given by 
\begin{eqnarray}
\Delta M^2 &=& 
\left \langle M^2 \right \rangle - \left ( \langle M \rangle  \right ) ^2 \nonumber \\
&=&
\frac{r_1 + r_2}{2} 
+ (-1)^{n_y + 1} \frac{ r_0 - r_1 }{2} \exp{\left (-D \right )}
\sin{ \Phi } \nonumber \\
&~&
+ \frac{ (r_0 - r_1)^2 }{4}
\left [
1 - \exp{ (-2D )} \sin^2{ \Phi}  
\right ].
\label{eq:SignalVariance}
\end{eqnarray}
Assuming that the detection noise is only caused by the photon shot noise characterized by the Poison distribution, 
the SNR is written as follows:
\begin{equation}
\label{eq:sqSNR}
\mathrm{SNR}^{-2} = \frac{1}{N_m} \frac{1}{N_{nv}}\frac{\Delta M^2}{(\Delta S)^2},
\end{equation}
where $N_{m}$ is the number of measurements
and $N_{nv}$ is the number of NV centers in an optical detection volume.
If we measure the phase shift of an AC magnetic field in $N\tau \sim T_2(N)$, 
the dominant terms in Eq. \eqref{eq:SignalVariance} are approximately given by 
\begin{equation}
\Delta M^2 \sim \frac{r_1 + r_2}{2}  + \frac{ (r_0 - r_1)^2 }{4}.
\end{equation}
Therefore, 
\begin{equation}
\label{eq:SNRapprox}
\mathrm{SNR} 
\sim \sqrt{N_{m} N_{nv}} C 
\exp \left \{ - \left [ \frac{N\tau}{T_2(N)} \right ]^{p} \right \}
\left | \cos{ \Phi } \frac{\partial \Phi }{ \partial \phi_{ac} } \right | |\Delta \phi_{ac}|,
\end{equation}
where $ C = \left [ 1 + 2(r_{0} + r_{1})/(r_{0} - r_{1} ) ^2 \right ] ^{-1/2}$.
If $\tau(1+\alpha) = \left ( 2f_{ac} \right ) ^{-1}$ and $\phi_{ac} = \pi/2$, 
Eq. \eqref{eq:SNRapprox} is consistent with Eq.~\eqref{eq:SNR}.

\subsection*{Data availability}
The data that support the findings of this study are available from the corresponding
author upon reasonable request.

%\bibliographystyle{naturemag}
%\bibliography{bib}	%Bibliography database file. Here bib.bib file is read.

\begin{thebibliography}{10}
\expandafter\ifx\csname url\endcsname\relax
  \def\url#1{\texttt{#1}}\fi
\expandafter\ifx\csname urlprefix\endcsname\relax\def\urlprefix{URL }\fi
\providecommand{\bibinfo}[2]{#2}
\providecommand{\eprint}[2][]{\url{#2}}

\bibitem{doi:10.1063/1.3491215}
\bibinfo{author}{Dang, H.~B.}, \bibinfo{author}{Maloof, A.~C.} \&
  \bibinfo{author}{Romalis, M.~V.}
\newblock \bibinfo{title}{Ultrahigh sensitivity magnetic field and
  magnetization measurements with an atomic magnetometer}.
\newblock \emph{\bibinfo{journal}{Appl. Phys. Lett.}}
  \textbf{\bibinfo{volume}{97}}, \bibinfo{pages}{151110}
  (\bibinfo{year}{2010}).

\bibitem{doi:10.1063/1.2943282}
\bibinfo{author}{Degen, C.~L.}
\newblock \bibinfo{title}{Scanning magnetic field microscope with a diamond
  single-spin sensor}.
\newblock \emph{\bibinfo{journal}{Appl. Phys. Lett.}}
  \textbf{\bibinfo{volume}{92}}, \bibinfo{pages}{243111}
  (\bibinfo{year}{2008}).

\bibitem{RevModPhys.81.1495}
\bibinfo{author}{Wiesendanger, R.}
\newblock \bibinfo{title}{Spin mapping at the nanoscale and atomic scale}.
\newblock \emph{\bibinfo{journal}{Rev. Mod. Phys.}}
  \textbf{\bibinfo{volume}{81}}, \bibinfo{pages}{1495--1550}
  (\bibinfo{year}{2009}).

\bibitem{JILES1988311}
\bibinfo{author}{Jiles, D.}
\newblock \bibinfo{title}{Review of magnetic methods for nondestructive
  evaluation}.
\newblock \emph{\bibinfo{journal}{NDT International}}
  \textbf{\bibinfo{volume}{21}}, \bibinfo{pages}{311 -- 319}
  (\bibinfo{year}{1988}).

\bibitem{Degen2008}
\bibinfo{author}{Degen, C.~L.}
\newblock \bibinfo{title}{Microscopy with single spins}.
\newblock \emph{\bibinfo{journal}{Nature Nanotechnology}}
  \textbf{\bibinfo{volume}{3}}, \bibinfo{pages}{643} (\bibinfo{year}{2008}).

\bibitem{Taylor2008}
\bibinfo{author}{Taylor, J.~M.} \emph{et~al.}
\newblock \bibinfo{title}{High-sensitivity diamond magnetometer with nanoscale
  resolution}.
\newblock \emph{\bibinfo{journal}{Nat. Phys.}} \textbf{\bibinfo{volume}{4}},
  \bibinfo{pages}{810} (\bibinfo{year}{2008}).

\bibitem{doi:10.1146/annurev-physchem-040513-103659}
\bibinfo{author}{Schirhagl, R.}, \bibinfo{author}{Chang, K.},
  \bibinfo{author}{Loretz, M.} \& \bibinfo{author}{Degen, C.~L.}
\newblock \bibinfo{title}{Nitrogen-vacancy centers in diamond: Nanoscale
  sensors for physics and biology}.
\newblock \emph{\bibinfo{journal}{Annu. Rev. Phys. Chem.}}
  \textbf{\bibinfo{volume}{65}}, \bibinfo{pages}{83--105}
  (\bibinfo{year}{2014}).

\bibitem{0034-4885-77-5-056503}
\bibinfo{author}{Rondin, L.} \emph{et~al.}
\newblock \bibinfo{title}{Magnetometry with nitrogen-vacancy defects in
  diamond}.
\newblock \emph{\bibinfo{journal}{Rep. Prog. Phys.}}
  \textbf{\bibinfo{volume}{77}}, \bibinfo{pages}{056503}
  (\bibinfo{year}{2014}).

\bibitem{Dolde2011}
\bibinfo{author}{Dolde, F.} \emph{et~al.}
\newblock \bibinfo{title}{Electric-field sensing using single diamond spins}.
\newblock \emph{\bibinfo{journal}{Nat. Phys.}} \textbf{\bibinfo{volume}{7}},
  \bibinfo{pages}{459} (\bibinfo{year}{2011}).

\bibitem{PhysRevA.95.053417}
\bibinfo{author}{Chen, E.~H.} \emph{et~al.}
\newblock \bibinfo{title}{High-sensitivity spin-based electrometry with an
  ensemble of nitrogen-vacancy centers in diamond}.
\newblock \emph{\bibinfo{journal}{Phys. Rev. A}} \textbf{\bibinfo{volume}{95}},
  \bibinfo{pages}{053417} (\bibinfo{year}{2017}).

\bibitem{Ariyaratne2018}
\bibinfo{author}{Ariyaratne, A.}, \bibinfo{author}{Bluvstein, D.},
  \bibinfo{author}{Myers, B.~A.} \& \bibinfo{author}{Jayich, A. C.~B.}
\newblock \bibinfo{title}{Nanoscale electrical conductivity imaging using a
  nitrogen-vacancy center in diamond}.
\newblock \emph{\bibinfo{journal}{Nat Commun}} \textbf{\bibinfo{volume}{9}},
  \bibinfo{pages}{2406} (\bibinfo{year}{2018}).

\bibitem{Kucsko2013}
\bibinfo{author}{Kucsko, G.} \emph{et~al.}
\newblock \bibinfo{title}{Nanometre-scale thermometry in a living cell}.
\newblock \emph{\bibinfo{journal}{Nature}} \textbf{\bibinfo{volume}{500}},
  \bibinfo{pages}{54} (\bibinfo{year}{2013}).

\bibitem{Toyli8417}
\bibinfo{author}{Toyli, D.~M.}, \bibinfo{author}{de~las Casas, C.~F.},
  \bibinfo{author}{Christle, D.~J.}, \bibinfo{author}{Dobrovitski, V.~V.} \&
  \bibinfo{author}{Awschalom, D.~D.}
\newblock \bibinfo{title}{Fluorescence thermometry enhanced by the quantum
  coherence of single spins in diamond}.
\newblock \emph{\bibinfo{journal}{Proc. Natl. Acad. Sci. U.S.A.}}
  \textbf{\bibinfo{volume}{110}}, \bibinfo{pages}{8417--8421}
  (\bibinfo{year}{2013}).

\bibitem{PhysRevB.91.155404}
\bibinfo{author}{Wang, J.} \emph{et~al.}
\newblock \bibinfo{title}{High-sensitivity temperature sensing using an
  implanted single nitrogen-vacancy center array in diamond}.
\newblock \emph{\bibinfo{journal}{Phys. Rev. B}} \textbf{\bibinfo{volume}{91}},
  \bibinfo{pages}{155404} (\bibinfo{year}{2015}).

\bibitem{Kolkowitz1603}
\bibinfo{author}{Kolkowitz, S.} \emph{et~al.}
\newblock \bibinfo{title}{Coherent sensing of a mechanical resonator with a
  single-spin qubit}.
\newblock \emph{\bibinfo{journal}{Science}} \textbf{\bibinfo{volume}{335}},
  \bibinfo{pages}{1603--1606} (\bibinfo{year}{2012}).

\bibitem{Balasubramanian2008}
\bibinfo{author}{Balasubramanian, G.} \emph{et~al.}
\newblock \bibinfo{title}{Nanoscale imaging magnetometry with diamond spins
  under ambient conditions}.
\newblock \emph{\bibinfo{journal}{Nature}} \textbf{\bibinfo{volume}{455}},
  \bibinfo{pages}{648} (\bibinfo{year}{2008}).

\bibitem{Rondin2013}
\bibinfo{author}{Rondin, L.} \emph{et~al.}
\newblock \bibinfo{title}{Stray-field imaging of magnetic vortices with a
  single diamond spin}.
\newblock \emph{\bibinfo{journal}{Nat. Commun.}} \textbf{\bibinfo{volume}{4}},
  \bibinfo{pages}{2279} (\bibinfo{year}{2013}).

\bibitem{Gross2017}
\bibinfo{author}{Gross, I.} \emph{et~al.}
\newblock \bibinfo{title}{Real-space imaging of non-collinear antiferromagnetic
  order with a single-spin magnetometer}.
\newblock \emph{\bibinfo{journal}{Nature}} \textbf{\bibinfo{volume}{549}},
  \bibinfo{pages}{252} (\bibinfo{year}{2017}).

\bibitem{Yane1603137}
\bibinfo{author}{Yan, S.} \emph{et~al.}
\newblock \bibinfo{title}{Nonlocally sensing the magnetic states of nanoscale
  antiferromagnets with an atomic spin sensor}.
\newblock \emph{\bibinfo{journal}{Sci. Adv.}} \textbf{\bibinfo{volume}{3}},
  \bibinfo{pages}{e1603137} (\bibinfo{year}{2017}).

\bibitem{Mamin557}
\bibinfo{author}{Mamin, H.~J.} \emph{et~al.}
\newblock \bibinfo{title}{Nanoscale nuclear magnetic resonance with a
  nitrogen-vacancy spin sensor}.
\newblock \emph{\bibinfo{journal}{Science}} \textbf{\bibinfo{volume}{339}},
  \bibinfo{pages}{557--560} (\bibinfo{year}{2013}).

\bibitem{Staudacher561}
\bibinfo{author}{Staudacher, T.} \emph{et~al.}
\newblock \bibinfo{title}{Nuclear magnetic resonance spectroscopy on a
  (5-nanometer)$^3$ sample volume}.
\newblock \emph{\bibinfo{journal}{Science}} \textbf{\bibinfo{volume}{339}},
  \bibinfo{pages}{561--563} (\bibinfo{year}{2013}).

\bibitem{doi:10.1021/nl402286v}
\bibinfo{author}{Ohashi, K.} \emph{et~al.}
\newblock \bibinfo{title}{Negatively charged nitrogen-vacancy centers in a 5 nm
  thin $^{12}$C diamond film}.
\newblock \emph{\bibinfo{journal}{Nano Lett.}} \textbf{\bibinfo{volume}{13}},
  \bibinfo{pages}{4733--4738} (\bibinfo{year}{2013}).

\bibitem{Lovchinsky503}
\bibinfo{author}{Lovchinsky, I.} \emph{et~al.}
\newblock \bibinfo{title}{Magnetic resonance spectroscopy of an atomically thin
  material using a single-spin qubit}.
\newblock \emph{\bibinfo{journal}{Science}} \textbf{\bibinfo{volume}{355}},
  \bibinfo{pages}{503--507} (\bibinfo{year}{2017}).

\bibitem{Aslam67}
\bibinfo{author}{Aslam, N.} \emph{et~al.}
\newblock \bibinfo{title}{Nanoscale nuclear magnetic resonance with chemical
  resolution}.
\newblock \emph{\bibinfo{journal}{Science}} \textbf{\bibinfo{volume}{357}},
  \bibinfo{pages}{67--71} (\bibinfo{year}{2017}).

\bibitem{Kotler2011}
\bibinfo{author}{Kotler, S.}, \bibinfo{author}{Akerman, N.},
  \bibinfo{author}{Glickman, Y.}, \bibinfo{author}{Keselman, A.} \&
  \bibinfo{author}{Ozeri, R.}
\newblock \bibinfo{title}{Single-ion quantum lock-in amplifier}.
\newblock \emph{\bibinfo{journal}{Nature}} \textbf{\bibinfo{volume}{473}},
  \bibinfo{pages}{61} (\bibinfo{year}{2011}).

\bibitem{Balasubramanian2009}
\bibinfo{author}{Balasubramanian, G.} \emph{et~al.}
\newblock \bibinfo{title}{Ultralong spin coherence time in isotopically
  engineered diamond}.
\newblock \emph{\bibinfo{journal}{Nat. Mater.}} \textbf{\bibinfo{volume}{8}},
  \bibinfo{pages}{383} (\bibinfo{year}{2009}).

\bibitem{1367-2630-13-4-045021}
\bibinfo{author}{Pham, L.~M.} \emph{et~al.}
\newblock \bibinfo{title}{Magnetic field imaging with nitrogen-vacancy
  ensembles}.
\newblock \emph{\bibinfo{journal}{New J. Phys.}} \textbf{\bibinfo{volume}{13}},
  \bibinfo{pages}{045021} (\bibinfo{year}{2011}).

\bibitem{GULLION1990479}
\bibinfo{author}{Gullion, T.}, \bibinfo{author}{Baker, D.~B.} \&
  \bibinfo{author}{Conradi, M.~S.}
\newblock \bibinfo{title}{New, compensated carr-purcell sequences}.
\newblock \emph{\bibinfo{journal}{J. Magn. Reson.}}
  \textbf{\bibinfo{volume}{89}}, \bibinfo{pages}{479 -- 484}
  (\bibinfo{year}{1990}).

\bibitem{PhysRev.94.630}
\bibinfo{author}{Carr, H.~Y.} \& \bibinfo{author}{Purcell, E.~M.}
\newblock \bibinfo{title}{Effects of diffusion on free precession in nuclear
  magnetic resonance experiments}.
\newblock \emph{\bibinfo{journal}{Phys. Rev.}} \textbf{\bibinfo{volume}{94}},
  \bibinfo{pages}{630--638} (\bibinfo{year}{1954}).

\bibitem{1807.00946}
\bibinfo{author}{Ishikawa, T.}, \bibinfo{author}{Yoshizwa, A.},
  \bibinfo{author}{Mawatari, Y.}, \bibinfo{author}{Kashiwaya, S.} \&
  \bibinfo{author}{Watanabe, H.}
\newblock \bibinfo{title}{Influence of dynamical decoupling sequences with
  finite-width pulses on quantum sensing for ac magnetometry}.
\newblock \bibinfo{note}{Preprint at} \eprint{https://arxiv.org/abs/1807.00946}
  (\bibinfo{year}{2018}).

\bibitem{doi:10.1063/1.1716296}
\bibinfo{author}{Meiboom, S.} \& \bibinfo{author}{Gill, D.}
\newblock \bibinfo{title}{Modified spin‐echo method for measuring nuclear
  relaxation times}.
\newblock \emph{\bibinfo{journal}{Rev. Sci. Instrum.}}
  \textbf{\bibinfo{volume}{29}}, \bibinfo{pages}{688--691}
  (\bibinfo{year}{1958}).

\bibitem{PhysRevX.5.041001}
\bibinfo{author}{Wolf, T.} \emph{et~al.}
\newblock \bibinfo{title}{Subpicotesla diamond magnetometry}.
\newblock \emph{\bibinfo{journal}{Phys. Rev. X}} \textbf{\bibinfo{volume}{5}},
  \bibinfo{pages}{041001} (\bibinfo{year}{2015}).

\bibitem{PhysRevLett.82.2417}
\bibinfo{author}{Viola, L.}, \bibinfo{author}{Knill, E.} \&
  \bibinfo{author}{Lloyd, S.}
\newblock \bibinfo{title}{Dynamical decoupling of open quantum systems}.
\newblock \emph{\bibinfo{journal}{Phys. Rev. Lett.}}
  \textbf{\bibinfo{volume}{82}}, \bibinfo{pages}{2417--2421}
  (\bibinfo{year}{1999}).

\bibitem{PhysRevB.77.174509}
\bibinfo{author}{Cywi\ifmmode~\acute{n}\else \'{n}\fi{}ski, L.},
  \bibinfo{author}{Lutchyn, R.~M.}, \bibinfo{author}{Nave, C.~P.} \&
  \bibinfo{author}{Das~Sarma, S.}
\newblock \bibinfo{title}{How to enhance dephasing time in superconducting
  qubits}.
\newblock \emph{\bibinfo{journal}{Phys. Rev. B}} \textbf{\bibinfo{volume}{77}},
  \bibinfo{pages}{174509} (\bibinfo{year}{2008}).

\bibitem{PhysRevA.79.062324}
\bibinfo{author}{Biercuk, M.~J.} \emph{et~al.}
\newblock \bibinfo{title}{Experimental Uhrig dynamical decoupling using trapped
  ions}.
\newblock \emph{\bibinfo{journal}{Phys. Rev. A}} \textbf{\bibinfo{volume}{79}},
  \bibinfo{pages}{062324} (\bibinfo{year}{2009}).

\bibitem{Bylander2011}
\bibinfo{author}{Bylander, J.} \emph{et~al.}
\newblock \bibinfo{title}{Noise spectroscopy through dynamical decoupling with
  a superconducting flux qubit}.
\newblock \emph{\bibinfo{journal}{Nature Physics}}
  \textbf{\bibinfo{volume}{7}}, \bibinfo{pages}{565} (\bibinfo{year}{2011}).
\newblock \bibinfo{note}{Article}.

\bibitem{Kawakami11738}
\bibinfo{author}{Kawakami, E.} \emph{et~al.}
\newblock \bibinfo{title}{Gate fidelity and coherence of an electron spin in an
  Si/SiGe quantum dot with micromagnet}.
\newblock \emph{\bibinfo{journal}{Proc. Natl. Acad. Sci. U.S.A.}}
  \textbf{\bibinfo{volume}{113}}, \bibinfo{pages}{11738--11743}
  (\bibinfo{year}{2016}).

\bibitem{cond-mat/0411174}
\bibinfo{author}{Devoret, M.~H.}, \bibinfo{author}{Wallraff, A.} \&
  \bibinfo{author}{Martinis, J.~M.}
\newblock \bibinfo{title}{Superconducting qubits: a short review}.
\newblock \bibinfo{note}{Preprint at} 
\eprint{https://arxiv.org/abs/cond-mat/0411174}
 (\bibinfo{year}{2004}).

\bibitem{0034-4885-80-10-106001}
\bibinfo{author}{Wendin, G.}
\newblock \bibinfo{title}{Quantum information processing with superconducting
  circuits: a review}.
\newblock \emph{\bibinfo{journal}{Rep. Prog. Phys.}}
  \textbf{\bibinfo{volume}{80}}, \bibinfo{pages}{106001}
  (\bibinfo{year}{2017}).

\bibitem{GU20171}
\bibinfo{author}{Gu, X.}, \bibinfo{author}{Kockum, A.~F.},
  \bibinfo{author}{Miranowicz, A.}, \bibinfo{author}{Liu, Y.} \&
  \bibinfo{author}{Nori, F.}
\newblock \bibinfo{title}{Microwave photonics with superconducting quantum
  circuits}.
\newblock \emph{\bibinfo{journal}{Phys. Rep.}}
  \textbf{\bibinfo{volume}{718-719}}, \bibinfo{pages}{1 -- 102}
  (\bibinfo{year}{2017}).

\bibitem{PhysRevLett.119.023602}
\bibinfo{author}{Kono, S.} \emph{et~al.}
\newblock \bibinfo{title}{Nonclassical photon number distribution in a
  superconducting cavity under a squeezed drive}.
\newblock \emph{\bibinfo{journal}{Phys. Rev. Lett.}}
  \textbf{\bibinfo{volume}{119}}, \bibinfo{pages}{023602}
  (\bibinfo{year}{2017}).

\bibitem{PhysRevLett.120.040505}
\bibinfo{author}{Eddins, A.} \emph{et~al.}
\newblock \bibinfo{title}{Stroboscopic qubit measurement with squeezed
  illumination}.
\newblock \emph{\bibinfo{journal}{Phys. Rev. Lett.}}
  \textbf{\bibinfo{volume}{120}}, \bibinfo{pages}{040505}
  (\bibinfo{year}{2018}).

\bibitem{PhysRevX.7.041011}
\bibinfo{author}{Bienfait, A.} \emph{et~al.}
\newblock \bibinfo{title}{Magnetic resonance with squeezed microwaves}.
\newblock \emph{\bibinfo{journal}{Phys. Rev. X}} \textbf{\bibinfo{volume}{7}},
  \bibinfo{pages}{041011} (\bibinfo{year}{2017}).

\bibitem{Clark2016}
\bibinfo{author}{Clark, J.~B.}, \bibinfo{author}{Lecocq, F.},
  \bibinfo{author}{Simmonds, R.~W.}, \bibinfo{author}{Aumentado, J.} \&
  \bibinfo{author}{Teufel, J.~D.}
\newblock \bibinfo{title}{Observation of strong radiation pressure forces from
  squeezed light on a mechanical oscillator}.
\newblock \emph{\bibinfo{journal}{Nat. Phys.}} \textbf{\bibinfo{volume}{12}},
  \bibinfo{pages}{683} (\bibinfo{year}{2016}).

\bibitem{RevModPhys.82.1155}
\bibinfo{author}{Clerk, A.~A.}, \bibinfo{author}{Devoret, M.~H.},
  \bibinfo{author}{Girvin, S.~M.}, \bibinfo{author}{Marquardt, F.} \&
  \bibinfo{author}{Schoelkopf, R.~J.}
\newblock \bibinfo{title}{Introduction to quantum noise, measurement, and
  amplification}.
\newblock \emph{\bibinfo{journal}{Rev. Mod. Phys.}}
  \textbf{\bibinfo{volume}{82}}, \bibinfo{pages}{1155--1208}
  (\bibinfo{year}{2010}).

\bibitem{7466817}
\bibinfo{author}{Watanabe, H.} \emph{et~al.}
\newblock \bibinfo{title}{Formation of nitrogen-vacancy centers in
  homoepitaxial diamond thin films grown via microwave plasma-assisted chemical
  vapor deposition}.
\newblock \emph{\bibinfo{journal}{IEEE Trans. Nanotechnol.}}
  \textbf{\bibinfo{volume}{15}}, \bibinfo{pages}{614--618}
  (\bibinfo{year}{2016}).

\bibitem{Bar-Gill2013}
\bibinfo{author}{Bar-Gill, N.}, \bibinfo{author}{Pham, L.~M.},
  \bibinfo{author}{Jarmola, A.}, \bibinfo{author}{Budker, D.} \&
  \bibinfo{author}{Walsworth, R.~L.}
\newblock \bibinfo{title}{Solid-state electronic spin coherence time
  approaching one second}.
\newblock \emph{\bibinfo{journal}{Nat. Commun.}} \textbf{\bibinfo{volume}{4}},
  \bibinfo{pages}{1743} (\bibinfo{year}{2013}).

\bibitem{doi:10.1063/1.4952418}
\bibinfo{author}{Sasaki, K.} \emph{et~al.}
\newblock \bibinfo{title}{Broadband, large-area microwave antenna for optically
  detected magnetic resonance of nitrogen-vacancy centers in diamond}.
\newblock \emph{\bibinfo{journal}{Rev. Sci. Instrum.}}
  \textbf{\bibinfo{volume}{87}}, \bibinfo{pages}{053904}
  (\bibinfo{year}{2016}).

\bibitem{doi:10.1063/1.3483676}
\bibinfo{author}{Meriles, C.~A.} \emph{et~al.}
\newblock \bibinfo{title}{Imaging mesoscopic nuclear spin noise with a diamond
  magnetometer}.
\newblock \emph{\bibinfo{journal}{J. Chem. Phys.}}
  \textbf{\bibinfo{volume}{133}}, \bibinfo{pages}{124105}
  (\bibinfo{year}{2010}).

\end{thebibliography}

\section*{Acknowledgements}
We acknowledge Hitoshi Sumiya from the Sumitomo Electric Industries for supplying the diamond substrate
and Yoshikiyo Toyosaki from the Correlated Electronics Group in AIST for the technical support provided in photolithography.
This work was supported in part by SENTAN.JST, 
Grant-in-Aid for Young Scientists (B) Grant Number JP17K14079,
and JSPS Kakenhi no. JP15H05853.

\section*{Author contributions statement}
T.I. conceived the idea, refined the experimental setup for AC magnetometry, 
carried out the experiments, analyzed the data, and wrote the manuscript.
H.W. prepared NV centers in the isotopically purified diamond film, and supervised the study.
A.Y., S.K., and Y.M. contributed to set up the optical and microwave systems
and discussed the study. 
All authors reviewed the manuscript. 

\section*{Additional information}
\noindent\textbf{Competing Interests:} The authors declare no competing interests.

\newpage

\newpage

\begin{table}[htbp]
	\caption{Evaluation of coherence time using Hahn-echo, CPMG-32, CPMG-128, and CPMG256.}
	\begin{center}
		\begin{tabular}{lccc}
			\hline \hline
			Pulse sequence & $N$ & $T_2(N) $ & $p$ \\
			\hline
			Hahn-echo & $1 $ & $ 74\pm3 ~ \mathrm{\mu s} $ & $0.95 \pm 0.04$ \\
			CPMG-32 & $32 $ & $ 340 \pm 10 ~ \mathrm{\mu s} $ & $1.3 \pm 0.1$ \\
			CPMG-128 & $128 $ & $ 650 \pm 20 ~ \mathrm{\mu s} $ & $1.2 \pm 0.1$ \\
			CPMG-256 & $256 $ & $ 1.2 \pm 0.1~ \mathrm{m s} $ & $1.7 \pm 0.4$ \\
			\hline \hline
		\end{tabular}
	\end{center}
	\label{tab:Fitting}
\end{table}

\begin{table}[htbp]
	\caption{Parameters of the CPMG-2, XY4-1, and XY8-1 sequences. 
				The listed parameters were used for fitting and theoretical plots,
				as explained in the main text.}
	\begin{center}
		\begin{tabular}{lcccc}
			\hline \hline
			Pulse sequence & $N$ & $T_2(N)$ ($\mathrm{\mu s}$) & $p$ & $r$\\
			\hline
			CPMG--2& $2$ & $95\pm4 $ & $1.11  \pm 0.06$    & $0.892 \pm 0.001 $ \\
			XY4--1 &   $4$ & $124 \pm 5   $ & $1.21  \pm 0.08$    & $0.908 \pm 0.001 $ \\
			XY8--1 &   $8$ & $140 \pm 10 $ & $0.97 \pm 0.09$ & $0.917 \pm 0.001 $ \\
			\hline \hline
		\end{tabular}
	\end{center}
	\label{tab:Parameters}
\end{table}

\end{document}